# Performance Enhancement via Real-time Image-based Beam Tracking for WA-OWC with Dynamic Waves and Mobile Receivers

Yujie DI, Anzi XU, and Lian-Kuan Chen, *Senior Member, IEEE*

*Abstract*—Intensified underwater activities have driven the escalating demand for reliable, flexible, and high data-rate underwater communication links. Optical wireless communication (OWC) emerges as the most promising technology for short- to medium-range communication, facilitating the real-time high-speed transmission of information from undersea to an aerial vehicle which can subsequently relay the information to a terrestrial station. However, establishing a robust water-air link confronts two primary challenges: (i) beam wandering due to the time-varying refraction when the light beam passes through the undulating ocean surface and (ii) the drone's movement when it hovers above the ocean surface. This paper experimentally demonstrated a real-time imaged-based beam tracking system to mitigate beam misalignment due to dynamic waves and receiver movement over a 0.14-m underwater and 1.83-m free-space OWC channel. Experimental results evince a notable reduction in the standard deviation of the received light spot offset from the receiver. Moreover, the tracking system can proficiently accommodate receiver velocities of up to 150 cm/s while maintaining a paltry packet loss rate (PLR) below 10%. By addressing the combined effects of dynamic waves and moving receivers, the proposed beam tracking system successfully enables a 70% reduction in PLR and an order of magnitude decrease in bit error rate (BER). This results in a substantial 17-fold surge in maximum throughput, from 50 Mbit/s to 850 Mbit/s. The experimental results validate the feasibility and effectiveness of the beam tracking system for vanquishing the detrimental effects in the complex water-air OWC (WA-OWC) channel and supporting high-speed data transmission.

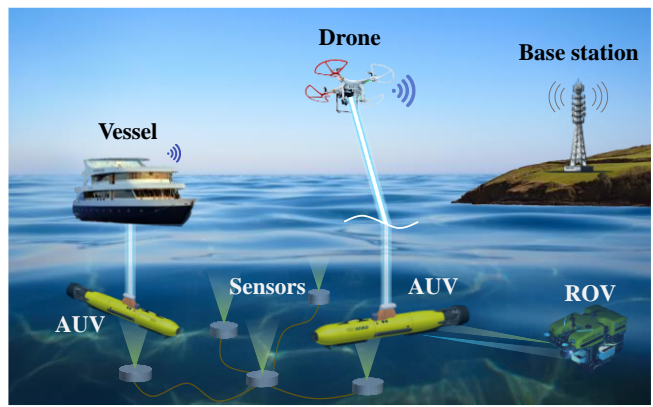

Fig. 1. Illustration of water-air OWC systems (ROV: remotely operated vehicles, AUV: autonomous underwater vehicle).

## I. INTRODUCTION

INTENSIFIED underwater activities, such as deep-sea mining, underwater rescue operations, and monitoring of oceanic ecosystems necessitate the development of efficient communication methods to transmit substantial amounts of undersea data to aerial vehicles and, subsequently, to terrestrial stations. The exigency for flexible and reliable communication links to support high data throughput becomes more pronounced. Traditional communication methodologies fall short in this scenario. Acoustic waves are hindered by their limited bandwidth, high latency, and significant reflection issues at the water surface, thus, rendering them ill-suited for high-speed applications [1]. Radiofrequency (RF) signals, commonly applied in wireless communication, suffer from severe attenuation in the water. On the other hand, cable-based communication, while stable, restricts the mobility of remotely operated vehicles (ROVs); their umbilical connections limit coverage, agility, and autonomy, especially in scenarios involving multiple ROVs with aggravated risk of cable entanglement. In contrast, optical wireless communication (OWC) is a promising solution for short- to medium-range communication scenarios, attributed to its high bandwidth, relative flexibility, and moderate loss in the water [2]. In [3], a real-time 2.2-Gbit/s time-multiplexed OWC system was realized for 4K video transmission over a 3.6-m underwater and an 8-m air optical wireless channel. By using a two-stage injection-locking technique, the modulation bandwidth of a laser diode (LD) could be significantly enhanced from 1.8 GHz to 18 GHz, realizing a 500-Gbit/s 100-m free-space and 10-m piped underwater link [4]. These studies have demonstrated the potential of high-speed water-air OWC (WA-OWC) systems

This Manuscript received xxx xx, 2023; revised xxx xx, 2023; accepted xxx xx, 2023. Date of publication xxx xx, 2023; date of current version xxx xx, 2023. This work was supported in part by the HKSAR RGC grant (GRF 14219322). (Corresponding author: Lian-Kuan Chen.)

Yujie Di, Anzi Xu, and Lian-Kuan Chen are with the Department of Information Engineering, The Chinese University of Hong Kong, Shatin, N. T., Hong Kong SAR. (e-mail: lkchen@ie.cuhk.edu.hk).

but were conducted with still water surfaces, ignoring the most challenging and practical impairment—the ever-changing water surface.

In real scenarios, WA-OWC encounters significant challenges, such as turbulence, bubbles, and waves [2]. Among these, wave presents a distinct and unique obstacle, one that is relatively unexplored in OWC research. When light passes through the water-air interface, the beam will be significantly deflected, leading to the laser beam offset from the receiver (Rx) and, consequently, a severe deterioration in communication performance. Several studies have explored effective mitigation methods to counteract the impact of wave-induced disturbances. A novel wave-aware adaptive loading scheme was proposed in [5]. It exploited the predicted wave slope to derive the expected received power and then adjusted the loaded bit accordingly, showing a 96.2% increment in throughput. In [6], a multi-input multi-output structure was applied to decrease the system outage probability caused by the wave effectively. A more straightforward wave mitigation approach is expanding the beam's divergence angle, thus attaining a larger light spot to accommodate the wave-induced beam wandering. Based on this, a diffused line-of-sight system was developed with a transmission field of view of 25° and achieved a 43.7-Mbit/s OWC over 0.3-m water and 0.6-m air channel under a water wave of 15-mm height [7]. However, a larger divergence angle implies lower power density, which is unsuitable for long-distance transmission. To overcome the wave effect while not sacrificing power efficiency, we can employ a beam steering system for collimated laser beams to compensate actively for wave-induced beam deflection.

Beam steering plays a pivotal role in the acquisition, tracking, and pointing (ATP) mechanisms [8] for beam alignment between a transmitter and receiver in an OWC system. ATP is crucial in mobile free-space optical communications (FSO), such as satellite/ground-to-satellite/ground [9] and vehicle-ground communication [10], which have been widely investigated. In [11], a bidirectional tracking scheme was developed, and it effectively tracked the signal beam wandering due to atmospheric turbulence and mechanical vibrations, supporting a 1.28-Tbit/s transmission capacity. In [12], a high-data-rate mobile-FSO was demonstrated at 40 cm/s with a 2.1-m transmission distance. The ATP mechanism has also been applied to underwater optical wireless communication (UOWC) between mobile vehicles. Various sensors were used to estimate the pointing error of optical communication between underwater mobile platforms, and the error was then minimized via an ATP system [13].

Despite all these valuable prior works in FSO and UOWC, the beam steering technology is rarely applied in wave mitigation for the WA-OWC system. In [14], a 4×4 ultrasonic sensor array and a micro-electromechanical system (MEMS) mirror were utilized for a coarse wave profile reconstruction and beam steering, respectively, achieving a 4.2-Mbit/s optical wireless link over a 0.33-m air distance. In [15], we developed a 3×3 photodiode (PD) array-based beam tracking scheme for wave mitigation, which employed a transmitter (Tx)-side beam tracking without feedback from the Rx. However, the resolution and tolerable offset range were restricted by the number of PDs. Above the wave, the drone's movement is also an issue, as it leads to beam misalignment between transmitters and receivers.

In this work, we experimentally demonstrate a real-time Tx-side, image-based beam tracking system for a WA-OWC system under the influence of dynamic waves and mobile receivers. We detail the optimization of tracking algorithms and hardware implementation. Then, we investigate the effects of dynamic waves and the moving receiver, individually and in combination. The system's performance is evaluated in terms of (i) the standard deviation of the received light spot offset from the receiver, (ii) bit error rate (BER), and (iii) packet loss rate (PLR) under different wave levels, moving speeds of the receiver, and data rates. Experimental results reveal that with the help of beam tracking, the PLR dramatically reduces from over 90% to less than 15%, and the throughput significantly enhances from 50 Mbit/s to 850 Mbit/s at an average wave slope changing rate (ASCR) of 0.0906 rad/s and Rx moving speed of 1 m/s, validating the effectiveness and feasibility of the proposed beam tracking system and the potential of the video transmission from water to air via OWC. This work is an extension of the work in Ref. [16], with notable additions and advancements including:

i. The image processing and the control algorithms are optimized for a responsive yet stable tracking system.
ii. Detailed characterization of wave effect is discussed. The relationship between wave slope changing rate (SCR) and light spot moving speed at the Rx side is analyzed theoretically and experimentally, linking the wave effect and the receiver's movement issue.
iii. The hardware is upgraded to compress the response time of the tracking system from 35 ms to 7 ms.
iv. Extensive experimental results are provided and discussed. With the optimized tracking system, the trackable moving speed of the receiver increases from 18 cm/s to 150 cm/s.

The rest of this paper is structured as follows. Section II gives the detailed principles of the proposed real-time image-based beam tracking system. In Section III, the experimental setup and hardware implementation of the water-air OWC system are illustrated. Section IV presents the experimental results and discussion. Finally, Section V gives the conclusion of this work.

## II. Image-based Beam Tracking Principles

### A. Relationship between wave effect and light spot movement at Rx sides

In [14], we assumed that a vertical incident light passes through a water surface with a time-varying wave equation $f(t,\vec{x})$. Fig. 1(a) in [15] shows that the wave-induced light spot displacement $d$ at time $t$ can be denoted as:

$$d(t) = h \cdot \tan[\beta(t) - \alpha(t)] \\ = h \cdot \tan[\arcsin(n' \cdot \sin \gamma(t)) - \gamma(t)], \quad (1)$$

where $h$ is the air distance from the water surface to the receiver, and $n' = n_{water}/n_{air}$. $n_{water}$ and $n_{air}$ are the refractive index for water and air, respectively. $\alpha$ and $\beta$ are the angle of incidence and angle of refraction, respectively, which can be represented by the wave slope angle $\gamma$ according to Snell's Law. The required tracking speed depends on the light spot moving speed, $d'(t)$, where $(\cdot)'$ denotes the differential operator. By differentiation with respect to time $t$, $d'(t)$ is expressed as:

$$d'(t) = \frac{h}{\cos^2(\arcsin(n' \cdot \sin \gamma(t)) - \gamma(t))} \cdot \left( \frac{n' \cdot \cos \gamma(t)}{\sqrt{1 - (n' \cdot \sin \gamma(t))^2}} - 1 \right) \cdot \gamma'(t). \quad (2)$$

$d'(t)$ is related to $\gamma(t)$ and proportional to $h$ and $\gamma'(t)$. We define $\gamma'(t)$ as wave slope changing rate (SCR) to characterize the wave dynamic motion.

*B. Tx-side Image-based Beam Tracking Principle*

A light beam from the underwater transmitter will pass through the water-air interface and incident onto the receiver mounted on a UAV hovering above the water surface. The schematic and flowchart of the image-based beam tracking system are illustrated in Fig. 2 (a) and (b). A light beam emitted from a LD is reflected by a MEMS mirror. When passing through the water-air interface, the light beam is deflected due to the dynamic waves. A receiver is mounted on a moving platform that moves back and forth on a rail to emulate the drone hovering over the water surface under the influence of strong wind. Both wave-induced beam deflection and drone's movement will result in a beam misalignment between Tx and Rx. A corner cube retroreflector is mounted on the Rx side to reflect part of the light beam back to the Tx side with an offset. The reflected beam, which serves as feedback, is antiparallel to the incident light. A camera located at the Tx side captures the reflected beam upon receiving a trigger signal from a microcontroller unit (MCU). The camera's exposure time is set as 1 ms when considering the frame rate and the minimum light intensity for camera detection. Shorter exposure times yield a more responsive tracking system but with reduced light intensity on the image, resulting in detection inaccuracy. The obtained light spot coordinates, acquired through image processing, are forwarded to the MCU. Tilting angles in x- and y-directions will be deduced via the control algorithm and conveyed to the MEMS mirror. Finally, the MEMS mirror will conduct beam steering to restore the light spot to the receiver. A 2-ms delay is added to ensure that the MEMS mirror will complete the tilting before the next image capturing to prevent motion blur of images. It is worth mentioning that the camera operates in pulsed trigger mode, which means the exposure will start on the arrival of a trigger pulse. Thus, the exposure process is under control, and the camera will not start exposure until the MEMS mirror completes tilting. The details of the experimental setup are given in Section III.

*C. Image Processing Algorithm*

To accurately obtain the coordinates of the light spot, we develop an image-processing algorithm that effectively distinguishes and pinpoints the target light spot. The diagram of this algorithm is illustrated in Fig. 3. Firstly, the image is converted to grayscale and resized to a 100×100-pixel resolution with linear interpolation. Then, an adaptive threshold function is utilized to isolate regions with intensity values surpassing 20% of the image's peak intensity, thereby effectively identifying the probable zones of light spots, as shown in Fig. 3(iii). The subsequent opening operation of morphology refines the detected light spot regions by eliminating minor artifacts and consolidating fragmented areas (Fig. 3(iv)). Finally, a blob detection algorithm is applied to locate the desired light spot center (Fig. 3(v)). Contiguous pixels with intensities within the threshold intensity range, which aggregate into a cluster surpassing a specific area size, are identified as the desired light spots as in [16]. By fine-tuning the area-size parameter and threshold intensity, the algorithm selectively identifies and locates blobs of interest, thus ensuring high accuracy of the target light spot coordinates.

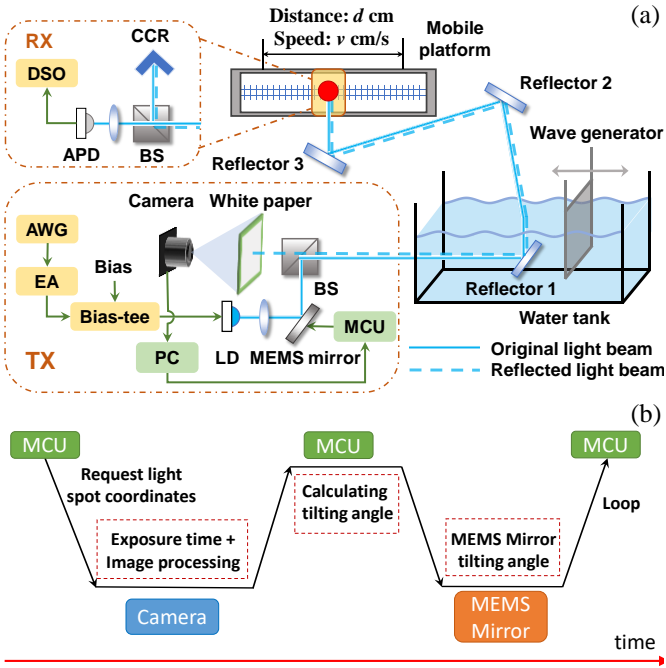

Fig. 2. (a) Experimental setup of real-time image-based beam tracking for WA-OWC system (AWG: arbitrary waveform generator; EA: electrical amplifier; LD: laser diode; MEMS mirror: micro-electromechanical system mirror; PC: personal computer; MCU: microcontroller unit; BS: beam splitter; APD: avalanche photodiode; CCR: corner cube reflector; DSO: digital storage oscilloscope.) and (b) block diagram of real-time image-based beam tracking system.

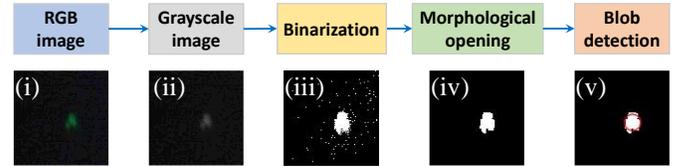

Fig. 3. Block diagram of image processing.

*D. Adaptive Control Algorithm*

Proportional–integral–derivative (PID) control provides rapid system responses and, thus, is particularly valuable for applications necessitating swift adjustments to adapt to dynamic environments [17]. Conventional PID control utilizes a fixed proportional coefficient ($k_p$) to compensate for discrepancies between the desired setpoint and output. The coefficient $k_p$ is crucial in determining the strength of the proportional control action. The selection of $k_p$ exhibits a trade-off: a larger $k_p$ yields a stronger and more responsive correction, albeit potentially leading to overshooting and oscillations. Conversely, a smaller $k_p$ provides a gentler correction, resulting in diminished responsiveness to changes.

The light spot offset at the Rx side fluctuates constantly due to the wave effect. Moreover, a drone's position and speed may continuously vary under the influence of the environment. Since the speed of the light spot is constantly changing, optimal $k_p$ is time-varying. Thus, a low-complexity adaptive real-time control, outlined in Algorithm 1 below, is developed to optimize the tracking performance. Only adjustments in the x-direction are shown, as the y-direction adjustment is similar. After camera capturing, the x coordinate of the light spot ($x_c$) can be derived via image processing. The light spot offset ($d_x$) from the target coordinate ($x_i$) is then calculated. $d_x^0$ represents the light spot offset in the previous frame and is initialized as 0.

The desired tilting angle will be derived via the PID control for beam steering to restore the beam spot to the Rx center. The proportional coefficient, $k_{px}$, is adjusted by multiplying $k_x$ with a scaling factor, $s_x$. A constant step size, $\alpha_x$, is employed to tune $s_x$. If the signs of $d_x^0$ and $d_x$ are opposite, and their absolute values are both greater than a constant, $p_1$, indicating potential oscillations due to overshooting, we decrease the value of $s_x$. Conversely, the proportional coefficient is probably too small if the signs of $d_x^0$ and $d_x$ are the same and their absolute values are larger than a constant, $p_2$. That hinders the prompt adjustment of the light spot back to the correct position; thus, $s_x$ should be increased. $k_p$ is initialized as the ratio of the tilting angle over the light spot offset. To avoid overshooting or undershooting, we confine $s_x$ to the interval [1, $q$].

---

**Algorithm 1.** Low-complexity Adaptive Control
**Input:** $s_x, k_x, \alpha_x, x_i, x_c, d_x^0, p_1, p_2, q$;
**Output:** $k_{px}$;

1. **while** true **do**
2.    $d_x = x_c - x_i$;
3.    **if** ($d_x \times d_x^0 < 0$) **then**
4.      **if** ($|d_x| > p_1$) and ($|d_x^0| > p_1$) **then**
5.        $s_x = s_x - \alpha_x$;
6.      **end if**
7.    **else if** ($|d_x| > p_2$) and ($|d_x^0| > p_2$) **then**
8.      $s_x = s_x + \alpha_x$;
9.    **end if**
10.   **if** $s_x < 1$ **then**
11.     $s_x = 1$;
12.   **else if** $s_x > q$ **then**
13.     $s_x = s_x - \alpha_x$;
14.   **end if**
15.   $k_{px} = s_x \times k_x$;
16.   $d_x^0 = d_x$;
17. **end while**

---

*E. Average Wave Slope Changing Rate (ASCR) Measurement*

In [14], we introduced the concept of the ASCR as an essential parameter for characterizing the rate of wave slope change. ASCR is highly related to communication performance that is influenced by the system's beam-tracking capability. Here, we will present the detailed characterization steps of ASCR for a comprehensive understanding and quantification of the wave motion.

A piece of paper is placed above and parallel to the water surface to serve as the receiver plane, the yellow rectangle shown in Fig. 4. A laser beam is directed perpendicularly onto the paper when there is no wave, generating a visible light spot. The dynamic behavior of the water surface is captured using a high-speed camera, operating at 220 frames per second (fps) to record 10,000 frames of the light spot wandering pattern. The light spot offset can be found for each frame. Then, by leveraging the exact correlation between the wave slopes and the light spot offset, we can deduce the wave slope and the ASCR.

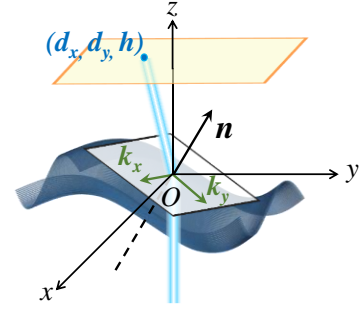

Fig. 4. The beam deflection and light spot offset induced by waves in 3D view.

Coordinates of the light spot are obtained via image processing introduced in Section II C. The slopes on the $xOz$ plane, $f_x$, and the $yOz$ plane, $f_y$, are then calculated by the light spot offset and air distance based on Snell's Law. As shown in Fig. 4, the tangent vector on the $xOz$ plane, $\mathbf{k_x} = (1, 0, f_x)$, and that on the $yOz$ plane, $\mathbf{k_y} = (0, 1, f_y)$, can be used to deduce the normal vector of the wave's tangent plane [18]:

$$\hat{\mathbf{n}} = \frac{\mathbf{k_x} \times \mathbf{k_y}}{\|\mathbf{k_x} \times \mathbf{k_y}\|} = \frac{1}{\sqrt{f_x^2 + f_y^2 + 1}}(-f_x, -f_y, 1). \quad (3)$$

The wave slope change (SC) between two adjacent frames (frame $i$ and $i+1$) is the angle between the two normal vectors, $\hat{\mathbf{n}}_i$ and $\hat{\mathbf{n}}_{i+1}$, of the wave's tangent planes in these two frames:

$$SC = \arccos(\hat{\mathbf{n}}_i \cdot \hat{\mathbf{n}}_{i+1}). \quad (4)$$

Then, the wave slope changing rate (SCR) is just SC/$\tau$ where $\tau$ is the corresponding time interval. The ASCR is computed as the mean of these SCRs over the entire observation period (=9,999 $\tau$)

III. EXPERIMENTAL SETUP AND IMPLEMENTATION

In this section, the experimental setup and the detailed hardware implementation of the real-time image-based WA-OWC system will be presented. Fig. 2 (b) shows the experimental setup of the proposed image-based beam tracking scheme for a WA-OWC system with a mobile receiver. A water tank with a dimension of 68 × 30 × 38 cm (length × width × depth) is filled with 0.14-m-deep tap water. As this study focuses on wave-induced impairment, water transmission distance is not the primary consideration. As for the air path, a longer air path will lead to a larger light beam offset and light spot moving speed, thereby escalating tracking difficulty. The air distance in this experiment is 183 cm. Waves are generated by a plastic board with reciprocating motion, and waves with varying levels can be generated by controlling the board's period, speed of movement, and travel distance. The corresponding ASCR and SC can be measured (detailed in Section II *E*) for each wave setting. The modulation signals are generated by an arbitrary waveform generator (AWG, Tektronix 7122 C) and amplified by an electrical amplifier (EA) of 12-dB gain. The amplified signal, coupled with a 6.0-V bias via a bias-tee (Mini-Circuits ZFBT-6GW+), is fed into a pigtail LD (495 nm, 33 mW). A MEMS mirror (Mirrorcle, A5L3.3-2400AL) is employed for beam steering with a diameter of 2.4 mm and a maximal tilting angle of around ±2.7 degrees for both x- and y-axes. The light is first focused by a lens and then reflected by the MEMS mirror. Considering our limited lab space and the challenge of positioning the mobile platform above the 1.83-meter-high water surface directly, three reflectors are strategically placed along the optical path. This configuration redirects the light beam from the water to the mobile platform, thereby extending the optical path. In real scenarios, those reflectors are not needed. Also, the laser beam under the water will be pointed upward to the water surface, and a drone will be above the water surface. In the experiment, the receiver is mounted on a mobile platform that is employed to emulate the drone's movement hovering above the water. The moving speed is adjustable, and the moving range is 20 cm. After passing through a 0.14-m water path and a 1.83-m air path, the light is split by a 50:50 beam splitter (BS) at the Rx side. Half of the light is reflected by a corner cube reflector (CCR) (Thorlabs, HRR201-P0) back to the Tx side. A camera (HTSUA33GC/M) behind a piece of paper captures images of the reflected light spot in trigger mode. A complete cycle of image capturing, light spot offset calculation, PID control, and beam steering takes around 7 ms, corresponding to a frame rate of 142.86 frames/s. At the Rx, the other half of the light is detected by a 1-GHz avalanche photodiode (APD) (Hamamatsu, C5658). The beam size at the Rx side is ~4 mm, and a lens is placed in front of the APD. The detected signal is recorded by a digital storage oscilloscope (DSO) for further offline signal processing. Each experiment collects 100 packets for analysis under different experimental conditions. A packet contains 10,000 symbols.

## IV. EXPERIMENTAL RESULTS AND DISCUSSION

In this section, the experimental results are presented to substantiate the effectiveness of the real-time image-based beam tracking system. Firstly, we evaluate the tracking system solely under the influence of dynamic waves while the receiver remains stationary. We characterize the wave and investigate the performance in terms of light spot fluctuation and communication quality under different wave conditions. Subsequently, we assess the system's performance with a moving receiver but no wave effect. Finally, we test the tracking system with the combined effects of both dynamic waves and receiver movement.

### A. Beam tracking under wave effect with stationary Rx

Fig. 5 (a) and (b) show the derived slope and SCR of the wave in the y-direction when the measured ASCR = 0.5155 rad/s. As the plastic board of the wave generator reciprocates in the y-direction, a distinct periodic curve can be observed. More than 10% of the SCRs have an absolute value exceeding 1 rad/s, and over 1% of |SCR| surpass 2 rad/s. The peak SCR value reaches up to 4.75 rad/s, posing a significant challenge to the tracking system. Fig. 5 (c) shows the recorded moving speed distribution of the light spot on the receiver plane, with over 27% of the moving speeds larger than 100 cm/s. These relatively high SCRs and light spot moving speeds could result in tracking failures due to the limited tilting speed of MEMS, thus deteriorating the communication performance.

Fig. 6 (a) and (b) display the trace of the light spot and the standard deviation of its displacement on the receiver plane, respectively. Apparently, a more concentrated trace of the light spot can be observed with tracking. As shown in Fig. 6 (b), the standard deviation along the y-axis is greater than that along the x-axis, attributed to the y-axis moving direction of the plastic board of the wave generator. Without tracking, the standard deviation on both the x- and y-axis increases with the increase of ASCR. In our wave generation, a wave with a larger ASCR also has a larger maximum wave slope, leading to a larger

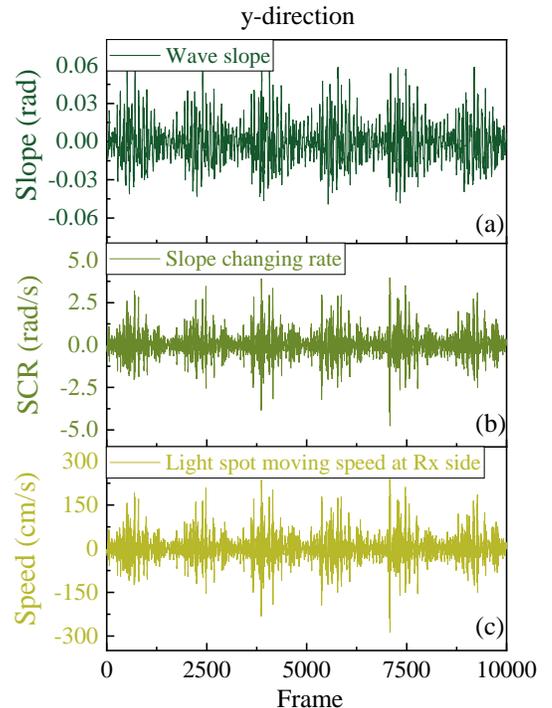

Fig. 5. Wave characterization at ASCR=0.5155 rad/s on y-direction for (a) slope, (b) slope changing rate, and (c) corresponding light spot moving speed at Rx side.

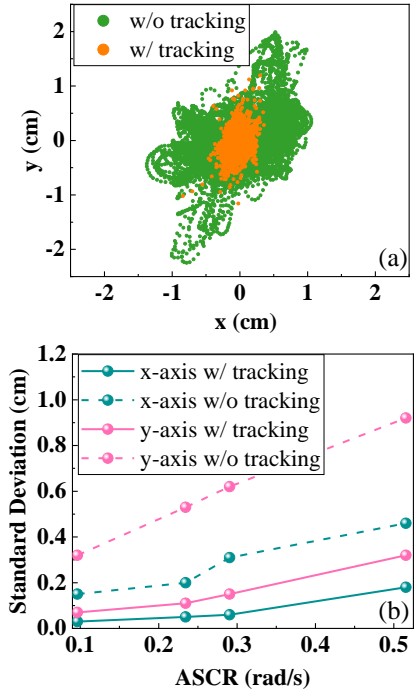

Fig. 6. (a) Light spot displacement on the receiver plane with ASCR=0.2344 rad/s and stationary Rx and (b) standard deviation of light spot displacement on the receiver plane versus wave ASCR

standard deviation of light spot offset in general. When tracking is enabled, as the ASCR escalates, the probability of light spot moving speeds surpassing the tracking speed also increases, resulting in performance deterioration.

At last, we investigate the performance of BER and PLR versus wave ASCR for 1-Gbit/s OOK signals with a stationary Rx. A packet is considered lost when the BER is higher than $3.8 \times 10^{-3}$, the hard-decision forward error correction (HD-FEC). Throughput is defined as the product of the data rate and the complement of the PLR, i.e., data rate * (1-PLR). Apparently, the OWC system with tracking outperforms that without tracking in all cases. As shown in Fig. 7, with tracking, the PLR remains below 20%, and notably, there is no packet loss at an ASCR of 0.0963 rad/s. Moreover, the throughput can achieve a notable 37% enhancement at ASCR of 0.5155 rad/s with beam tracking. Both BER and PLR performances deteriorate as ASCR increases, due to the increasing tracking failures. It is worth noting that the illustrated BER represents the average of 100 packets, and the BER value may be dominated by the worst

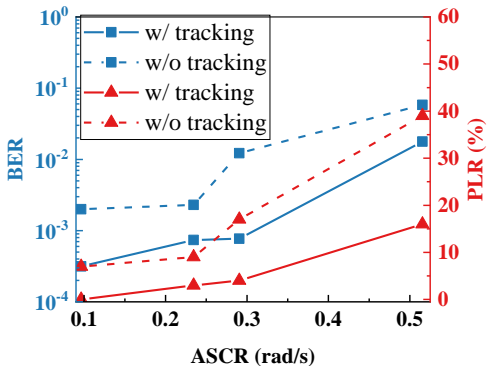

Fig. 7. BER and PLR performance versus wave ASCR for 1-Gbit/s OOK signals with stationary Rx.

packets.

*B. Beam tracking with moving Rx and no wave introduced*

We next evaluate the effectiveness of beam tracking on a mobile receiver platform in the absence of wave effect. The initial step is the BER measurement when the receiver is parked at a specific position on the moving rail (Fig. 8). With the tracking system activated, the light beam from Tx is precisely aligned with the Rx, leading to a zero BER for a stationary Rx at all tested positions. Conversely, without tracking, the BER surges as the Rx moves out of the light beam, as shown in Fig. 8. When the Rx terminal is positioned within a range of -7 mm to 7 mm for the 4-mm light spot, the BER can be reduced to below $3.8 \times 10^{-3}$, thanks to the coupling lens at APD. Thus, the tracking system needs to steer the beam to follow the moving Rx within ±7 mm.

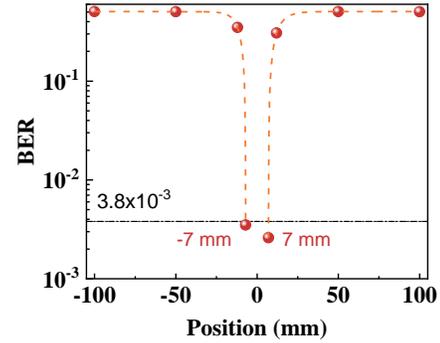

Fig. 8. BER when Rx parks at different positions on the rail without tracking and the beam is directed to the moving rail's center.

The light spot offset from the moving Rx and its standard deviation are recorded and displayed in Fig. 9 (a) and (b), respectively. With beam tracking, this offset can be reduced greatly from ±10 cm to ±0.7 cm in the x-direction. Moreover, the standard deviation on the x-axis decreases significantly, from over 7 cm to below 0.4 cm. The y-axis standard deviation is small for both with and without tracking, as the Rx is moving in the x-direction only.

We further investigate the BER and PLR performance of the beam tracking versus Rx moving speed for 1-Gbit/s OOK signals when no wave is introduced. As shown in Fig. 10, a significant reduction in PLR, from more than 90% to under 8%, is achieved, emphatically demonstrating the effectiveness of tracking. With the tracking enabled at 150-cm/s Rx moving speed, the throughput can reach 930 Mbit/s, a 31-fold increase from a mere 30 Mbit/s when there is no tracking. Furthermore, the implementation of the tracking scheme results in an improvement of more than an order of magnitude in BER.

*C. Beam tracking under wave effect with moving Rx*

At last, we evaluate the communication performance of the system versus the OOK signal's data rate, considering the combined effects of dynamic waves and the movement of the Rx terminal. The wave effect with an ASCR of 0.096 rad/s is introduced, along with a 100-cm/s Rx moving speed. As displayed in Fig. 11, the PLR and BER performances with tracking markedly outperform those without tracking. The PLR remains below 15% for data rates up to 1000 Mbit/s when tracking is employed. On the other hand, PLR dramatically surges to above 90% without tracking. Furthermore, the maximum throughput reaches 850 Mbit/s with tracking compared to only 50 Mbit/s without tracking. This significant enhancement in communication performance confirms that the proposed beam tracking not only effectively mitigates the wave-induced deterioration but also compensates for the beam misalignment due to the Rx terminal's movement. Other than wave-induced effects, in real-world WA-OWC scenarios, ocean surface layers may also exhibit other issues, such as the presence of bubbles, that may degrade tracking performance and need to be considered. The influence of bubbles on the ATP system was investigated in [19], revealing that the ATP system can mitigate bubble effects but may require prolonged pointing time. These bubbles may also induce dispersion, resulting in inter-symbol interference (ISI) and consequently restricting transmission distance and data rate. The ISI issue may be alleviated by signal processing, such as the adaptive RLS equalizer presented in [20].

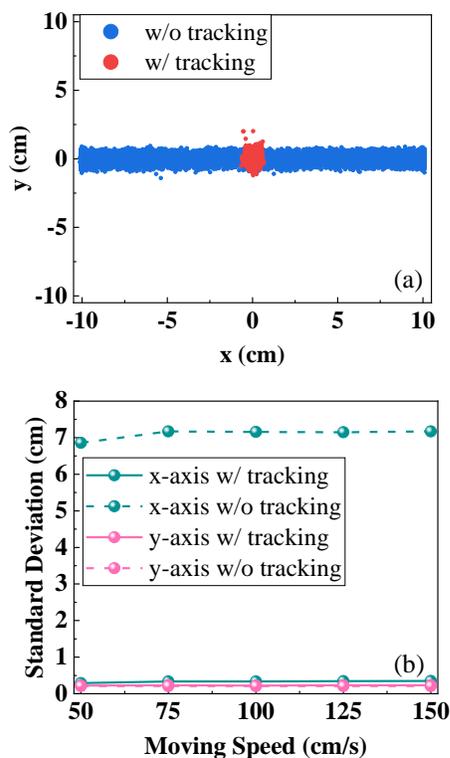

Fig. 9. (a) Light spot offset from the receiver center with 1-m/s moving speed and no wave introduced and (b) standard deviation of light spot offset from the receiver center versus moving speed of Rx.

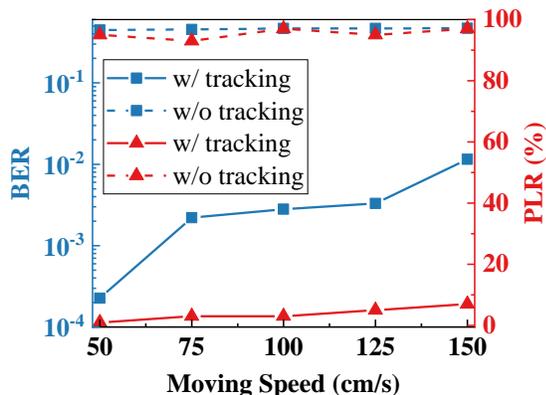

Fig. 10. BER and PLR performance versus moving speed of Rx for 1-Gbit/s OOK signals without wave effect.

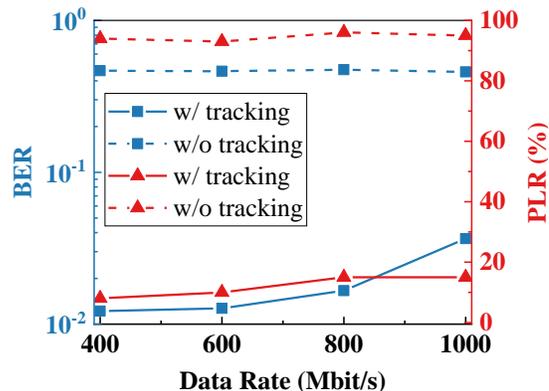

Fig. 11. BER and PLR performance versus data rate of OOK signals with a wave ASCR of 0.096 rad/s and Rx moving speed of 100 cm/s.

## V. Conclusion

In this paper, a real-time imaged-based beam tracking system is experimentally demonstrated in the presence of dynamic wave and Rx movement, conducted over a 0.14-m underwater and 1.83-m free-space OWC channel. The study provides a systematical and comprehensive investigation of the proposed beam tracking system for WA-OWC. The correlation between the wave ASCR and the received light spot moving speed is analyzed theoretically and experimentally. The standard deviation analysis is employed to examine the deviation of the received light spot from the center of the receiver. The results demonstrate that beam tracking leads to a much more stable light spot at the receiving end. Moreover, we extensively study the communication performances under the individual and combined influences of dynamic waves and moving receivers. The efficacy of the beam tracking system has been showcased by a pronounced decrement in both PLR and BER, with the latter diminished by an order of magnitude and the former curtailed by over 70%. Additionally, the maximum throughput is markedly enhanced to 850 Mbit/s, a seventeen-fold improvement, revealing the potential for high-speed data transmission within the WA-OWC system. The successful demonstration verifies the feasibility and effectiveness of the beam tracking system for wave mitigation and beam alignment between Tx and moving Rx terminals. This achievement paves the way for the practical deployment of WA-OWC systems, enabling real-time surveillance of underwater activities such as underwater mining and environmental monitoring, which are of great importance. In our future work, we will explore novel precoding-based approaches [21] to further optimize the performance of tracking-assisted WA-OWC systems. We aim to develop a novel channel-agnostic solution to attain a longer transmission distance and higher data rate.

## References

[1] Z. Zeng, S. Fu, H. Zhang, Y. Dong, and J. Cheng, "A survey of underwater optical wireless communications," in *IEEE Communications Surveys & Tutorials*, vol. 19, no. 1, pp. 204-238, 2017.

[2] L. K. Chen, Y. Shao, and Y. Di, "Underwater and Water Air Optical Wireless Communication," in *J. Light. Technol.*, vol. 40, no. 5, pp. 1440-1452, 2022.

[3] Y. Shao, R. Deng, J. He, K. Wu, and L. K. Chen, "Real-time 2.2-Gb/s Water-air OFDM OWC System with Low-complexity Transmitter-side DSP," in *J. Light. Technol.*, vol. 38, no. 20, pp. 5668–5675, 2020.

[4] W. Tsai *et al.*, "500 Gb/s PAM4 FSO-UWOC Convergent System With a R/G/B Five Wavelength Polarization-Multiplexing Scheme," in *IEEE Access*, vol. 8, pp. 16913–16921, 2020.

[5] Y. Shao, Y. Di, and L. K. Chen, "Adaptive loading for water-air SIMO OWC system based on the temporal and spatial properties of waves," in *Proc. OFC*, 2021, paper Th3E.2.

[6] Lin *et al.*, "Waving Effect Characterization for Water-to-Air Optical Wireless Communication," in *J. Light. Technol.*, vol. 41, no. 1, pp. 120-136, 2023.

[7] X. Sun, M. Kong, C. Shen, C. H. Kang, T. K. Ng, and B. S. Ooi, "On the realization of across wavy water-air-interface diffuse-line-of-sight communication based on an ultraviolet emitter," in *Opt. Express*, 27, pp. 19635–19649, 2019.

[8] Y. Kaymak, R. Rojas-Cessa, J. Feng, N. Ansari, M. Zhou, and T. Zhang, "A Survey on Acquisition, Tracking, and Pointing Mechanisms for Mobile Free-Space Optical Communications," in *IEEE Communications Surveys & Tutorials*, vol. 20, no. 2, pp. 1104-1123, 2018.

[9] V. W. S. Chan, "Optical satellite networks," in *J. Lightw. Technol.*, vol. 21, no. 11, pp. 2811–2827, Nov. 2003.

[10] S. Haruyama *et al.*, "New ground-to-train high-speed free-space optical communication system with fast handover mechanism," in *Proc. Opt. Fiber Commun. Conf.*, Los Angeles, CA, USA, 2011, pp. 1–3.

[11] E. Ciaramella *et al.*, "1.28 terabit/s (32x40 Gbit/s) wdm transmission system for free space optical communications," in *IEEE Journal on Selected Areas in Communications*, vol. 27, no. 9, pp. 1639-1645, 2009.

[12] Toshimasa Umezawa *et al.*, "Multi-Stacked Large-Aperture High-Speed PIN-Photodetector for Mobile-FSO Communication," in *49th European Conference on Optical Communication (ECOC 2023)*, Glasgow, Scotland, 2023, pp. 159 – 162.

[13] Y. Weng, T. Matsuda, Y. Sekimori, J. Pajarinen, J. Peters and T. Maki, "Pointing Error Control of Underwater Wireless Optical Communication on Mobile Platform," in *IEEE Photonics Technology Letters*, vol. 34, no. 13, pp. 699-702, 2022.

[14] C. J. Carver, Z. Tian, H. Zhang, K. M. Odame, A. Q. Li, and X. Zhou, "Amphilight: Direct air-water communication with laser light," in *17th USENIX NSDI 20*, 2020.

[15] Y. Di, Y. Shao, and L. K. Chen, "Real-Time Wave Mitigation for Water-Air OWC Systems via Beam Tracking," in *IEEE Photonics Technology Letters*, vol. 34, no. 1, pp. 47-50, 2022.

[16] Y. Di, A. Xu, and L. K. Chen, "Real-time Image-based Beam Tracking for Water-air OWC System with Mobile Receiver through Wavy Water Surface," in *49th European Conference on Optical Communication (ECOC 2023)*, Glasgow, Scotland, 2023, pp. 936 – 939.

[17] M. A. Johnson and M. H. Moradi, "PID control," London, UK: Springer-Verlag London Limited, 2005.

[18] Y. Tian and S. G. Narasimhan, "The relationship between water depth and distortion function," 2009.

[19] J. Lin *et al.*, "Machine-vision-based acquisition, pointing, and tracking system for underwater wireless optical communications," in *Chin. Opt. Lett.*, vol.19, no. 5, pp. 050604-, 2021.

[20] Y. Wang, X. Huang, L. Tao, J. Shi, and N. Chi, "4.5-Gb/s RGB-LED based WDM visible light communication system employing CAP modulation and RLS based adaptive equalization," in *Opt. Express*, vol. 23, no.10, pp. 13626-13633, 2015.

[21] G. Wang, D. Zhang, J. Zhao, and L. K Chen, "A unified precoding-based FTN scheme for communication channels with limited bandwidth," *49th European Conference on Optical Communication (ECOC 2023)*, Glasgow, Scotland, 2023, pp. 1095 – 1098.